\documentclass[superscriptaddress,twocolumn,showpacs,amssymb]{revtex4-1}

\usepackage[dvips]{graphicx}
\usepackage{latexsym}
\usepackage{graphicx}
\usepackage{float}
\usepackage[utf8]{inputenc}
\usepackage[english]{babel}
\usepackage{amsmath}
\usepackage{mathtools}
\usepackage{amssymb}
\usepackage{verbatim}
\usepackage{amsfonts}
\usepackage{hyperref}

\hypersetup{
	colorlinks=true,
	pdfborder={0 0 0},
}
\usepackage{color}
\usepackage[export]{adjustbox}
\usepackage{verbatim}
\newcommand{\XC}[1]{{\color{uv} #1}}

\definecolor{blue}{rgb}{0,0,1}
\definecolor{green}{rgb}{0,.6,0}
\definecolor{red}{rgb}{1,0,0}
\definecolor{vio}{rgb}{1,0,1}
\definecolor{uv}{rgb}{0.,0,0.}
\definecolor{ama}{rgb}{0.3,0.3,0.3}
\newcommand{\dif}{\mathrm{d}}

\newcommand{\im}{\mathbf{i}}

\newcommand{\abs}[1]{\left\vert#1\right\vert}

\begin{document}
\title{Incomplete thermalization from trap-induced integrability breaking: \\ lessons from classical hard rods}
\begin{abstract}
We study a one-dimensional gas of hard rods trapped in a harmonic potential, which breaks integrability of the hard-rod interaction in a non-uniform way.  We explore the consequences of such broken integrability for the dynamics of a large number of particles and find three distinct regimes: initial, chaotic, and stationary.  The initial regime is captured by an evolution equation for the phase-space distribution function.  For any finite number of particles, this hydrodynamics breaks down and the dynamics become chaotic after a characteristic time scale determined by the inter-particle distance and scattering length. \XC{The system fails to thermalize over the time-scale studied ($10^4$ natural units)}, but the time-averaged ensemble is a stationary state of the hydrodynamic evolution.  We close by discussing logical extensions of the results to similar systems of quantum particles.
\end{abstract}

\author{Xiangyu Cao}
\affiliation{Department of Physics, University of California, Berkeley, Berkeley CA 94720, USA}
\author{Vir B. Bulchandani}
\affiliation{Department of Physics, University of California, Berkeley, Berkeley CA 94720, USA}
\author{Joel E. Moore}
\affiliation{Department of Physics, University of California, Berkeley, Berkeley CA 94720, USA}
\affiliation{Materials Science Division, Lawrence Berkeley National Laboratory, Berkeley CA 94720, USA}
\date{\today}
\maketitle

\paragraph{Introduction.} It has been known since the work of Poincar{\'e} that even the simplest mechanical systems can exhibit complex dynamics, with chaotic behavior as the norm and integrability as a somewhat special case. This distinction is only sharpened as the number of degrees of freedom increases.  The time-evolution of a generic interacting many-body system is chaotic and ergodic: starting from any initial condition, trajectories of the system sample uniformly all configurations allowed by a few conservation laws, and are subject to the laws of statistical mechanics. The integrable many-body systems are exceptions to this rule, and are able to escape ergodicity and conventional thermalization thanks to the existence of an extensive number of conserved quantities.

In practice, exact integrability is fine-tuned and vulnerable to real-world imperfections, so that systems with broken integrability are more abundant than perfectly integrable ones. Moreover, broken integrability provides valuable insights into the general theory of dynamical systems. For example, in classical mechanics, the KAM theorem~\cite{Chierchia2010KAM} states that for weak enough perturbations, integrability is preserved in some finite portion of the phase space. For a uniform perturbation of a many-body system, the integrability-preserving phase-space often becomes vanishingly small, and no such ``gray zone'' is allowed. Such a sharp distinction extends in general to quantum many-body systems~\cite{GGE1},  although thermalization can be parametrically slow with weak integrability breaking~\cite{Essler14,Bertini15,Bertini16Thermalization}. 

In this work, we examine the consequences of the \textit{non-uniform} integrability breaking that results from placing an integrable many-body system in a trap. A famous experimental realization of this scenario is the ``quantum Newton's cradle'', which consists of a trapped, quasi one-dimensional Bose gas in a harmonic trap~\cite{QNC,tang2017thermalization}.  In the absence of the trap, the system is integrable and does not thermalize; even with the trap, which destroys the higher conservation laws needed for integrability, it is found that the system fails to thermalize over experimentally accessible time-scales. Here, the trap plays a delicate role: it is needed to observe periodic motion, rather than a simple expansion of the trapped gas, but also liable to destroy it eventually due to its breaking integrability. This raises two natural questions: what is the time scale $t_*$ induced by integrability breaking, defined as the advent of chaos, and does the system reach thermal equilibrium in the long-time limit?

In the present work, we address these questions by studying a \textit{classical} analogue, the one-dimensional gas of hard rods~\cite{Tonks} in a harmonic trap, whose time evolution can be obtained exactly from molecular dynamics simulations.  We propose the following simple scaling law for $t_*$, in terms of the \XC{\textit{constant}} potential curvature $V''(x)$, particle mass $m$, the scattering length (rod length) $a$ and the maximal gas density $\rho_m$:
\begin{equation}\label{eq:timetochaos}
1/t_*  = C  \rho_m a  \omega \,,\,  \omega =\sqrt{ V''(x) / m } ,
\end{equation}
where $C$ is an order-unity dimensionless pre-factor. Our short answer to the second question is: the system is chaotic but complete thermalization is \textit{not} observed in the long time scale accessible to us ($t \omega \sim 10^4$).  The full answer is quite elaborate and related to the other theme of this work: the validity of classical and quantum hydrodynamical equations in systems whose integrability is destroyed by a trap~\footnote{Note that integrability can be preserved in a trap, e.g., for the Calogero model, whose hydrodynamics has been also studied~\cite{Abanov05,Kulakani17}}.

We find that both the initial time regime and the long-time stationary ensemble are usefully captured by the kinetic theory of hard rods~\cite{Percus,Boldrighini,DoyonSpohn17}, while there is an intervening chaotic regime in which hydrodynamics fails.  An area of recent progress is that kinetic equations of the same (dissipationless Boltzmann) type capture the large-scale dynamics of {\it quantum} integrable systems~\cite{CastroAlvaredo16,Bertini16,DoyonYoshimura17,Ilievski17,BVKM1,BVKM2,Drude17,Alba17,Piroli17,Hubbard17}, with a self-consistent velocity functional drawn from the Bethe equations.  In the presence of a trap, the kinetic equation admits an extension~\cite{DoyonYoshimura17}, \XC{which has not been tested against microscopic dynamics. We perform this test in the context of the classical hard-rod gas}, as it is straightforward to write down a trapped hard-rod equation (tHRE) in the presence of an external potential. A direct comparison against microscopic simulations shows that the tHRE is accurate in an initial regime $t < t_*$, before breaking down for any finite system-size, due to a ``complexity crisis'' that will be explained. Despite the onset of chaos, we find that the late-time non-thermal ensembles are described by \textit{stationary} solutions to the tHRE~\cite{DoyonYoshimura17}.

\paragraph{Trapped hard-rod gas.} The hard-rod gas in a harmonic trap is equivalent to $N$ one-dimensional harmonic oscillators with hard-core repulsive interaction. The Hamiltonian reads
\begin{subequations}\label{eq:H}
	\begin{align}
	&H =  \sum_{j=1}^N \left[ \frac{1}{2}  p_j^2 + V(x_j) \right] +\sum_{j < k} U(x_j-x_k)  \\
	&V(x) = \frac12 \omega^2 x^2 \,,\,  U(\delta x) = \begin{cases}
	0 & \abs{\delta x}  > a \\ \infty & \abs{\delta x}  \leq a  \,,
	\end{cases}
	\end{align}
\end{subequations}
where $a > 0$ denotes the rod length, and $x_j$ and $p_j$ denote positions and momenta (we set $m = 1$). Upon re-scaling time as $t \to t \omega$, we may set $\omega = 1$ without loss of generality. Starting from a configuration such that $x_{j+1} - x_j \geq a$, $j=1, \dots, N-1$, the gas evolves as $N$ decoupled oscillators, until the next collision (\textit{i.e.}, $x_{j+1} - x_j = a$ for some $j$) in which the rods $j$ and $j+1$ exchange their velocities spontaneously. Such a dynamics can be efficiently and exactly simulated.
There are two integrable limits. Upon removing the trap, one recovers the usual hard-rod gas. Its momentum distribution is conserved and its dynamics map to those of $N$ independent particles. Meanwhile, in the limit of vanishing rod length $a = 0$, we obtain $N$ decoupled harmonic oscillators. Yet, in the presence of both trap and interaction, we find no other conserved quantities besides the total energy and the center-of-mass energy which we set to $0$~\footnote{The center of mass decouples from the relative coordinates and behaves as a simple harmonic oscillator.}.

\begin{figure}
	\center 
	\includegraphics[valign = t,width=.7\columnwidth]{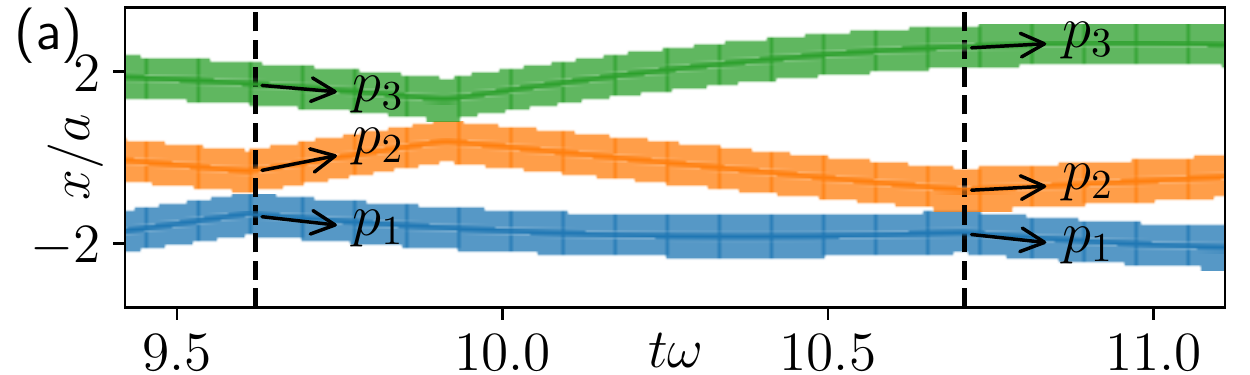} \\
	\includegraphics[valign = t,width=.7\columnwidth]{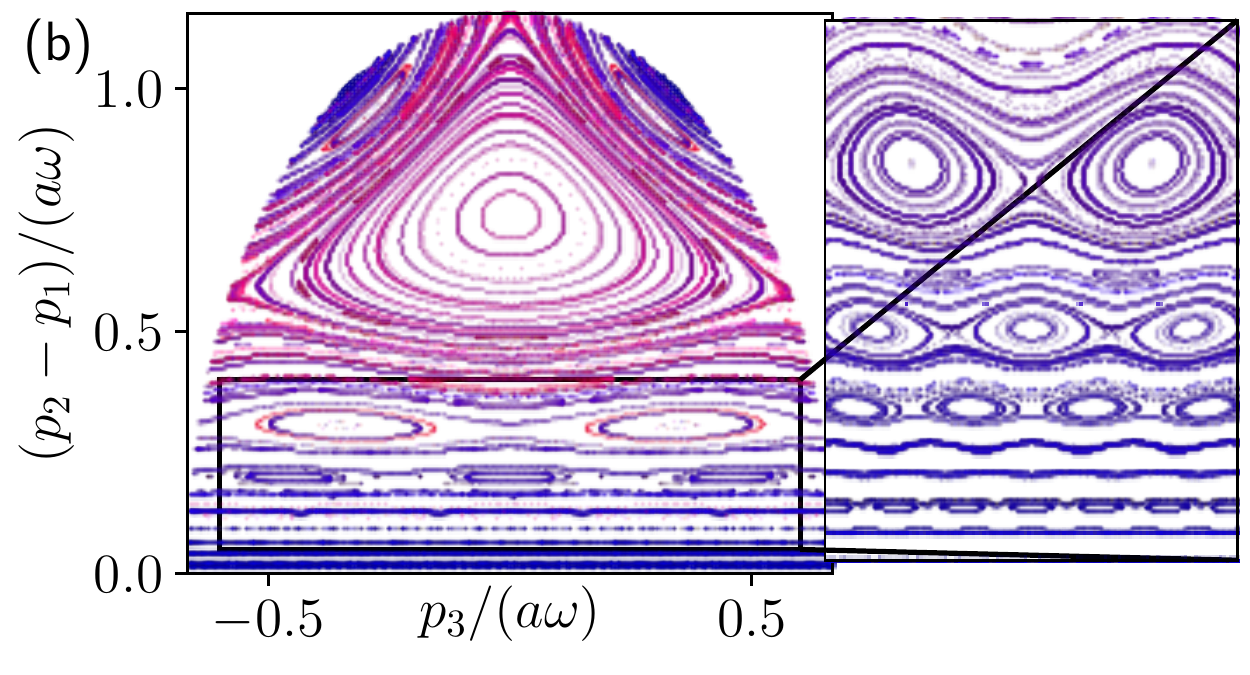} 
	\caption{(a) An illustration of three-rod dynamics. The Poincar\'e sector is defined as the set of configurations just after a $1$-$2$ collision. They are indicated by dashed lines. The Poincar\'e map sends the left one to the right one. (b)  Orbits of the  Poincar\'e recurrence map, with $H = 4$ and vanishing center-of-mass energy. The sector is bijectively parametrized by $p_2-p_1$ and $p_3$. Different colors distinguish distinct orbits. } \label{fig:poincare}
\end{figure}

To provide more convincing evidence of microscopic non-integrability, we studied the three-body problem. Its phase space, constrained by the conserved quantities, is three-dimensional and one can visualize the orbits of the Poincar\'e recurrence map, defined on a 2D sector of colliding configurations, as in Fig.~\ref{fig:poincare}. The fractal structure observed is inconsistent with the existence of any higher analytic integrals of motion. Yet, most trajectories do not cover the available phase space, so are not micro-canonical. 

\paragraph{Hydrodynamics.} The large-scale, coarse-grained dynamics of the hard-rod gas without the trap is described by a Boltzmann-type equation, which governs the single-particle phase space distribution $\rho(x,p) = \frac{\dif^2 N}{\dif x \dif p}$. Collisions conserve particles' momenta but modify their effective velocities. The resulting kinetic equation,
	\begin{align}
	&\partial_t \rho + \partial_x (v \rho) = 0,  v[\rho](p) = p +  \frac{a \int_{p'} (p-p') \rho(x,p') }{1 - a \int_{p'} \rho(x,p')}   \label{eq:boltzmann}
	\end{align}
was first obtained by Percus~\cite{Percus}, and rigorously proven~\cite{Boldrighini} to define an Euler-scale hydrodynamics of the hard-rod gas. Recently, equations similar to eq.~\eqref{eq:boltzmann} were shown to capture a variety of large-scale dynamics in quantum integrable systems~\cite{CastroAlvaredo16,Bertini16,DoyonYoshimura17,Ilievski17,BVKM1,BVKM2,Drude17,Alba17,Piroli17,Hubbard17}, in which context we call eq.~\eqref{eq:boltzmann} the Bethe-Boltzmann equation (BBE), since the analogue of $v[\rho](p)$ is obtained from thermodynamic Bethe ansatz. A modification of BBE in an external potential was proposed in ~\cite{DoyonYoshimura17}, which coincides with the standard Boltzmann correction for the Lieb-Liniger and quantum hard-rod models \cite{IntMB}. For classical hard-rods, the same correction can be obtained by different arguments~\footnote{This equation follows straightforwardly from the derivation given by Percus in ~\cite{Percus}, provided one assumes that the pair correlation function of the gas is not modified by the trapping potential at length scales of the order of a rod length, $a$.}, and yields
\begin{equation}
\partial_t \rho + \partial_x (v \rho) - \partial_x V \partial_p \rho = 0 \,. \label{eq:boltzmannV}
\end{equation}
Since the trap breaks integrability of the microscopic dynamics, the validity of eq.~\eqref{eq:boltzmannV} is so far a hypothesis to be tested.
 
Nevertheless, the tHRE is conceptually helpful as a guide to defining the thermodynamic ($N \to \infty$) limit. Indeed, eq.~\eqref{eq:boltzmannV} has an \textit{emergent} scale-invariance, $(\rho, a) \mapsto (\lambda \rho, a / \lambda)$ in any potential $V$, which relates pairs of systems with different $N$. Now, in a harmonic trap, we can further apply a spatial rescaling $(x, a) \to (\lambda x, \lambda a)$, and define profiles of different $N$ corresponding to a fixed hydrodynamic profile $\tilde{\rho}$, with $a$ fixed:
\begin{equation}
\rho(x,k) :=  \tilde\rho(\tilde{x} = x / N,  \tilde{p} = p / N)  / N \,.
\end{equation}
Therefore, we will set $a = 1$ in what follows.

We consider initial conditions (ICs) with Gaussian profiles:
$\tilde{\rho}(\tilde{x} ,\tilde{p}) =   \exp\left(-\frac{\tilde{x}^2}{2 \sigma_x^2} - \frac{\tilde{p}^2}{2\sigma_p^2} \right) / (2 \pi \sigma_p \sigma_x)$.
We can check that $\sigma = \sqrt{\sigma_x^2 + \sigma_p^2}$ and $N$ fixes the total energy. We also define a characteristic density:
\begin{equation}\label{eq:rhom}
\rho_m = 1/ \sqrt{ \pi \sigma } \,.
\end{equation}
$\rho_m$ is proportional to the density of $\rho(x,p)$ at origin, with a pre-factor depending only on $\sigma / \sigma_x$, which describes how ``squeezed'' the IC is. 

The results will be discussed in three consecutive time regimes: initial, chaotic and late-time. 

\paragraph{Initial regime and tHRE breakdown.} 
\begin{figure}
	\includegraphics[valign = t,width=.7\columnwidth]{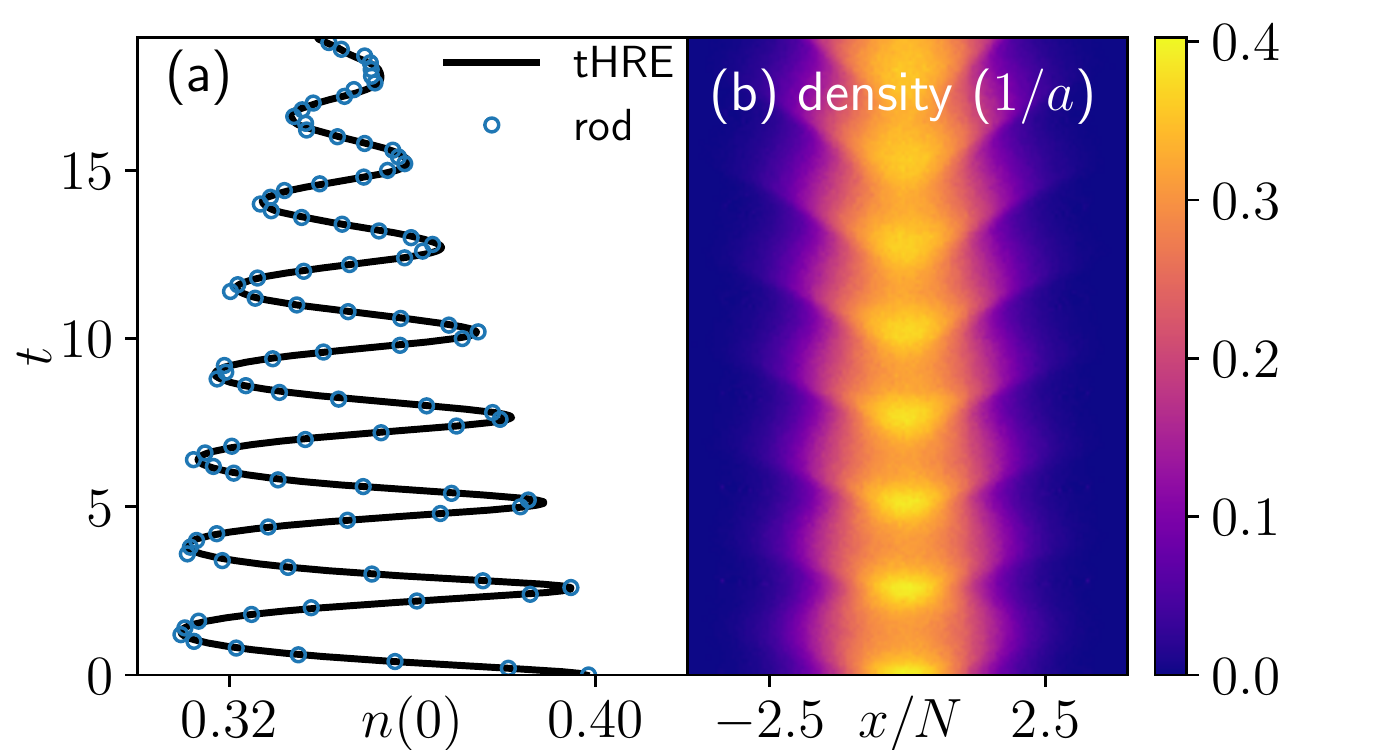}
	\caption{Comparing hard rod dynamics and tHRE in the initial regime. (a) Comparing the density at origin ($n(0)$) The hard rod data (circles) is obtained by averaging over $200$ realizations with  $N = 1024$ rods, representing a circular IC profile with $\sigma_x = \sigma_p$, $\rho_m a = 0.4$ [eq.~\eqref{eq:rhom}]. The tHRE data (curve) is obtained from its numerical integration~\cite{BVKM1,BVKM2}. (b): the density profile evolution obtained from the rod simulation.} \label{fig:compare}
\end{figure}
Using the protocols defined above, we can compare the exact microscopic dynamics against predictions from tHRE, which we expect to be valid at least at short times, when integrability remains unbroken. To this end, we adapt the scheme developed in Refs.~\cite{BVKM1,BVKM2} to solve numerically the tHRE; the microscopic result is averaged over many ICs sampling the same initial hydrodynamic profile. An example comparison is illustrated in Fig.~\ref{fig:compare}. A damped density oscillation is observed in the short-time dynamics, and is accurately captured by tHRE. Therefore, damping \textit{per se} is not a signature of integrability breaking.
To understand the nature of damping, we visualize the evolution of $\rho(x,p)$ in Fig.~\ref{fig:S}(a). Recall that the absence of interaction ($a = 0$) would lead to a simple rotation of the IC. \XC{The interaction induces a \textit{many-body dephasing} responsible for the damping. In the $x$-$p$ phase space, the dephasing generates a complex structure, reminiscent of a growing galaxy. Such galaxy formation is also observed in the numerical solution of the tHRE. Because the tHRE is dissipationless, we believe that the tHRE solution has ever-increasing complexity, which any finite-$N$ system cannot reproduce exactly: then, tHRE must break down, due to a ``complexity crisis''. Systems with larger $N$  have higher resolution and resist the complexity crisis better ~\footnote{\XC{Note that an anharmonic trap would induce a single-particle dephasing~\cite{caux2017hydrodynamics}, which also generates a complex phase space distribution, but is unrelated to integrability breaking and chaos. See Supplementary for more discussion.}}.} 

\XC{A quantification of the breakdown of tHRE is \textit{entropy} growth.} Indeed, the dissipationless tHRE conserves the entropy functional of the hard-rod gas,
\begin{equation}\label{eq:S}
S := \int_{x,p}  \rho \ln \theta,  \text{ where } \theta(x,p) := \frac{\rho(x,p)}{1 - a \int_{p'}\rho(x,p')} \,.
\end{equation}
Hence, measuring the time-evolution of $S$ from microscopic simulations tests the validity of tHRE without solving it directly. The result, in Fig.~\ref{fig:S}-b, shows a clear entropy growth after $t \sim 10$, invalidating tHRE at long time. The growth is suppressed for larger $N$, as expected. \XC{We also compared it to the estimate using the entropy production term known for a non-trapped hard-rod gas in local (generalized Gibbs) equilibrium~\cite{Boldrighini1997,DoyonSpohn17}. We find a qualitative agreement, which deteriorates quantitatively as $N$ increases. This suggests that the integrability breaking leads to some local equilibration (required by the entropy production term), which is suppressed when $N$ increases. We will support this scenario by studying dynamical chaos.} 

\begin{figure}	
	\begin{center}
	\includegraphics[valign = t,width=.7\columnwidth]{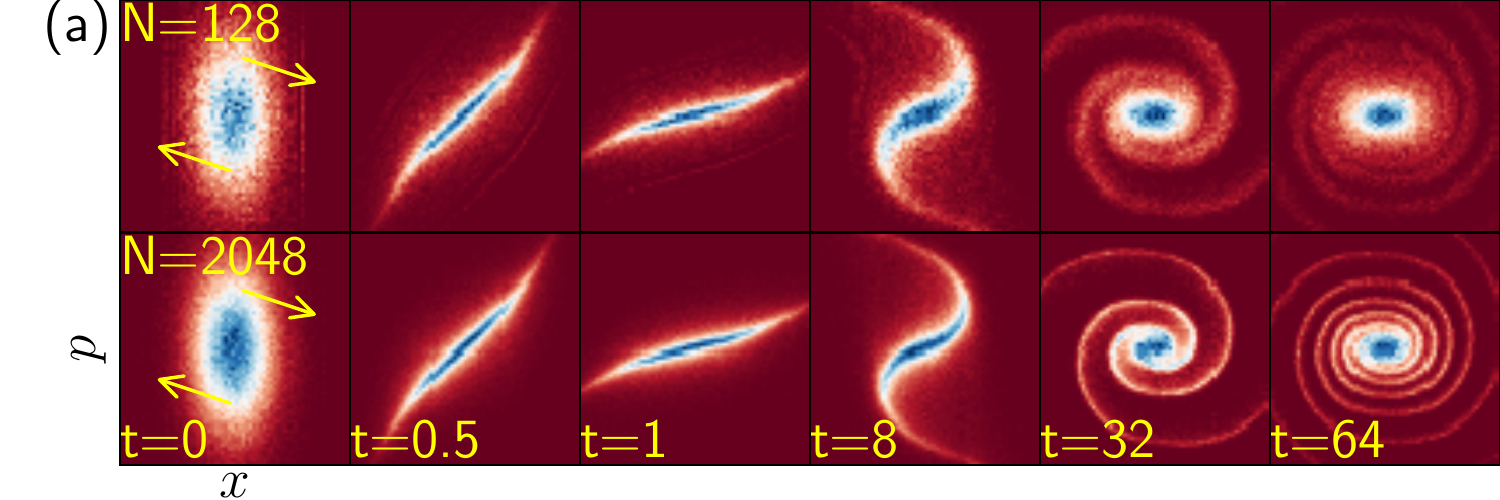} \\
	\includegraphics[valign = t,width=.7\columnwidth]{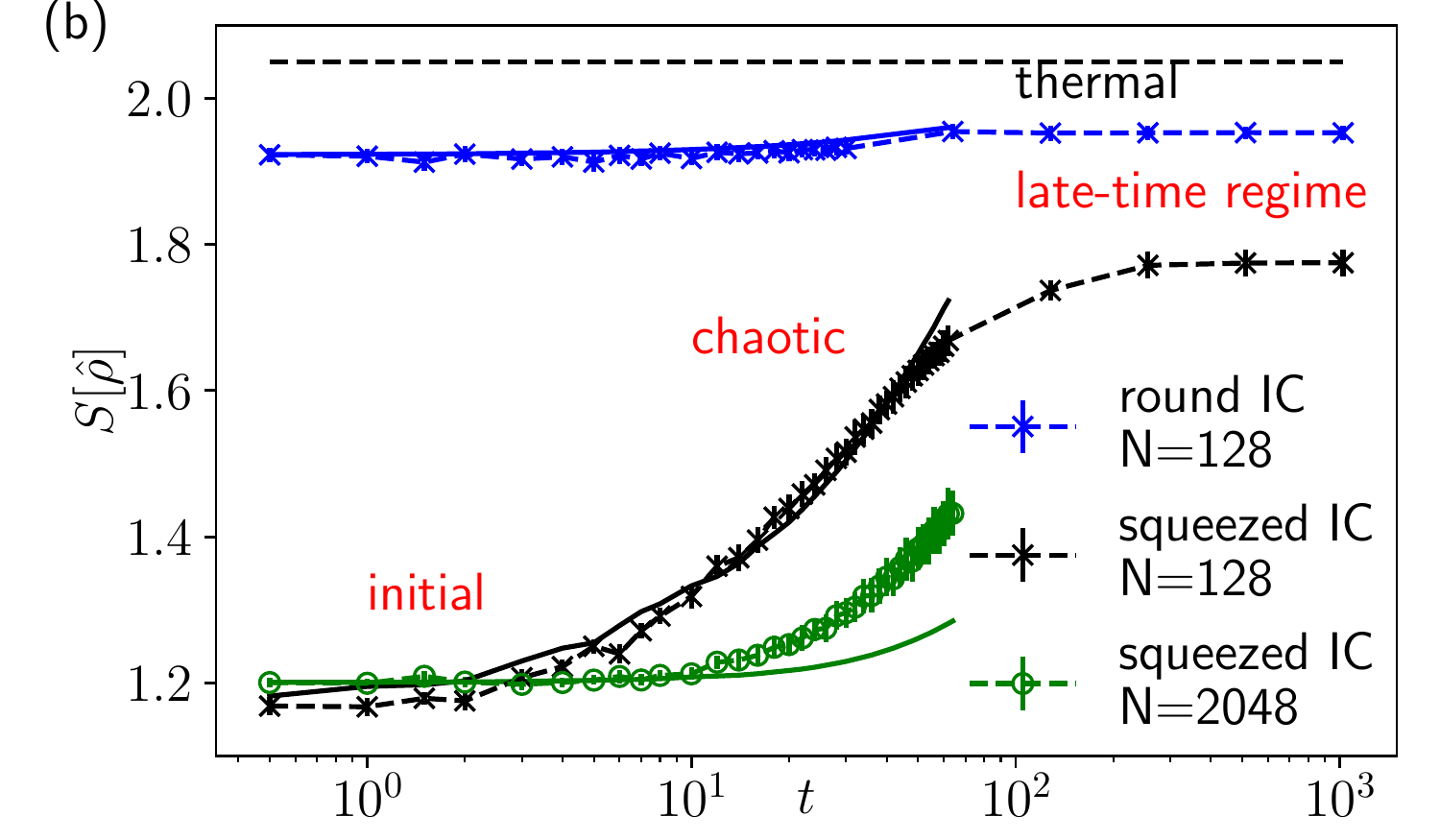} 
	\end{center}
	\caption{(a) Evolution of phase-space distribution $\rho(x,p)$ for the squeezed IC ($\rho_m = 1/2$, $\sigma_x = 1/2$), with different $N$. At $t = 64$, the $N = 2048$ system preserves a complex structure, which is completely smeared out for $N = 128$. (b) Entropy increase during time evolution, starting from different ICs. \XC{The solid curves represent entropy growth estimated using the entropy production term in ~\cite{DoyonSpohn17}}.
    } \label{fig:S}
\end{figure}

\begin{figure}
	\includegraphics[valign = t,width=.98\columnwidth]{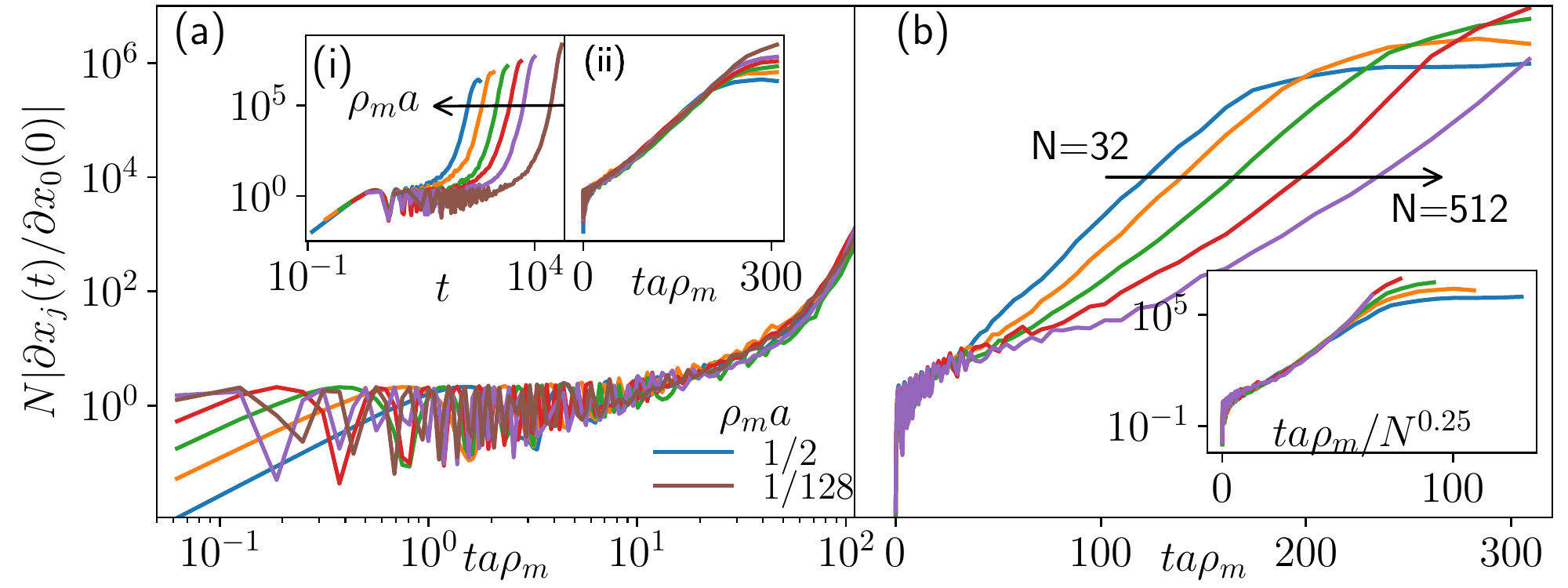}
	\caption{Dynamical chaos measured as separation of trajectories. (a)  Demonstrating the scaling law of time to chaos, eq.~\eqref{eq:timetochaos}. $N=64$ for all data. Main plot: Scaling collapse, initial regime and the crossover to chaos. (i) Raw data. (ii) Scaling collapse of the exponential chaos regime.  \XC{The initial-regime oscillation has a period independent of $\rho_m a$,  and close to that of a single harmonic oscillator.} (b)  Suppression of the chaos in larger systems, as illustrated by the decrease of the Lyapunov exponent. Main: raw data for $\rho_m a = 1/2$; inset: data collapse suggesting the scaling $\gamma   \propto  N^{-0.25}$.}\label{fig:chao}
\end{figure}

\paragraph{Advent of chaos.} \label{sec:chaos}
We measure chaos defined as the exponential separation of $N$-body phase space trajectories~\cite{Gutzwiller}. 
We first consider circular ICs with varying $\rho_m a$, apply small perturbations and measure the average deviation induced particle positions $\delta x_j(t)$ after evolution~\footnote{The perturbation is applied on the velocity of the leftmost rod. Other initial perturbations are considered, yielding the same scaling law. We do not include velocities for being discontinuous.}.  
The result, shown Fig.~\ref{fig:chao}-a, displays a clear cross-over from non-chaotic plateau ($|\delta x_j(t)| \sim |\delta x_1(0)|$) to a chaotic regime, at a time $1/t_* = C \rho_m a$, eq.~\eqref{eq:timetochaos}, where $C \approx 0.1$. \XC{We show in the Supplementary Material that the same time scaling law governs many-body dephasing and complexity crisis}. The data collapse in the chaos regime implies the Lyapunov exponent scaling $ \gamma \propto a \rho_m.$ The proportionality constants depend on other aspects of the IC, but is in general positively correlated with entropy growth. In particular, chaos is suppressed as $N$ increases. Indeed, our data are consistent with the power law
$ \gamma   \propto  N^{-v} \,, v \approx 0.25 \,,$ 
see Fig.~\eqref{fig:chao}-b.
Such a many-body suppression of chaos underlies that of local equilibration discussed above, and is probably a general feature of  ``weak'' integrability breaking, i.e., not by interactions, but by a trap. Although microscopic integrability is broken, infinitely many conserved quantities (including the entropy) of the HRE remain conserved for the tHRE \footnote{For example, both the HRE and the tHRE possess infinitely many conservation laws of the form $\int_{x,p}\rho f(\theta)$, where $f:\mathbb{R} \to\mathbb{R}$ is arbitrary, even though tHRE is not obviously integrable in the sense of ~\cite{VB17}.}, and only affected by higher-derivative terms to tHRE, which are finite-$N$ corrections. 

\paragraph{Late-time ensemble.}\label{sec:hydro2}
The previous results all involve averaging over some IC ensemble. From now on, we focus on single, long trajectories, and average only over time. In a generic thermalizing system, such a time-averaged ensemble converges swiftly to the microcanonical ensemble, even when the IC is highly atypical thermodynamically (consider filling only a half of a box with gas). Thermalization is usually associated with chaos, since exponential separation of nearby trajectories means that the initial condition is quickly forgotten. Therefore, one would expect that the trapped hard-rod gas thermalizes at late times $t \gg t_*$. 

\begin{figure}
	\center
	\includegraphics[valign = t,width=.49\columnwidth]{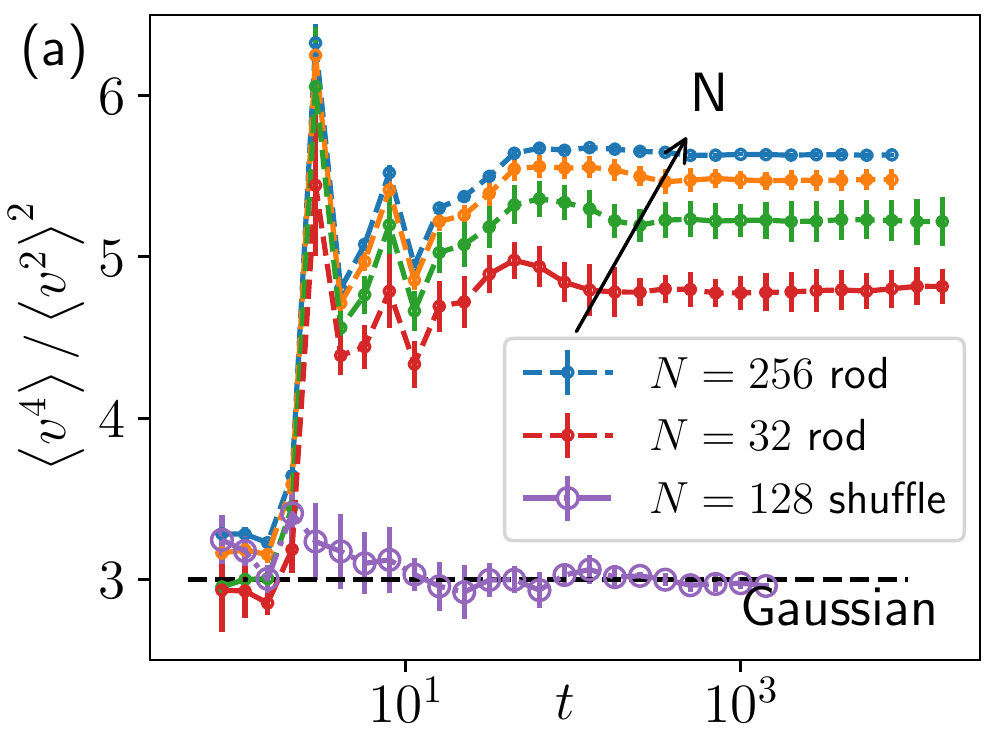}
	\includegraphics[valign = t,width=.49\columnwidth]{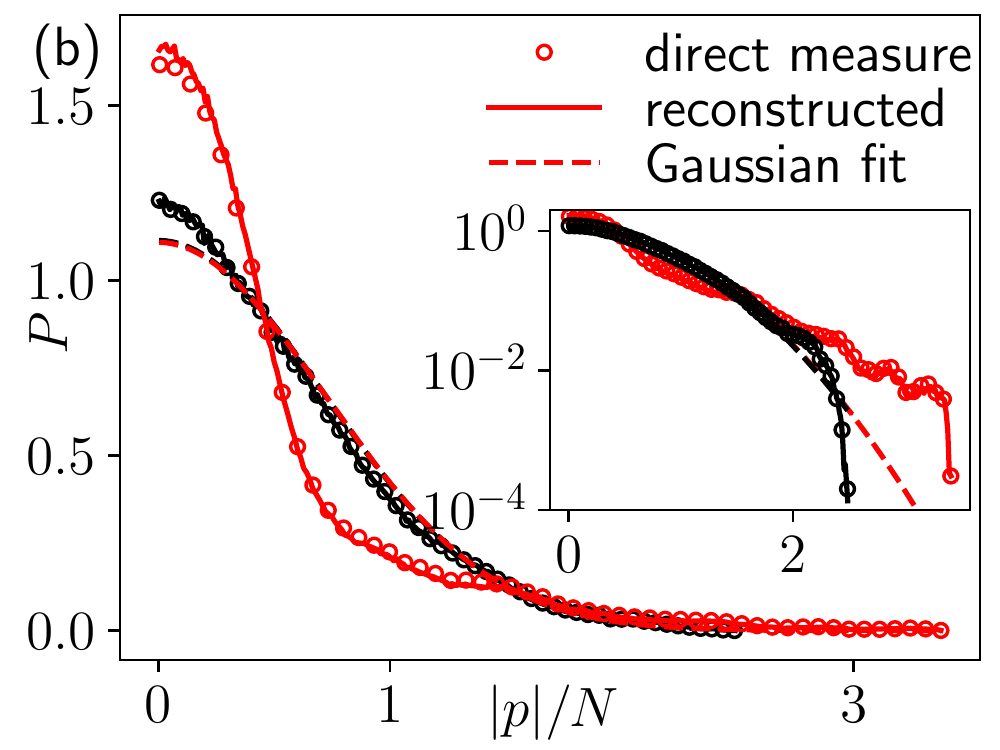}
	\caption{(a) (Non)-Gaussianity of the velocity distribution of the time-averaged ensemble, as revealed by the moment-ratio test. (b) Comparing the velocity distribution with the reconstructed one from the density, assuming that the late time ensemble solves the stationary tHRE eq.~\eqref{eq:HRE0}. The two long-time ensembles are obtained from two squeezed (red) and circular (black) ICs, both with $\rho_m a = 1/2$. }\label{fig:late}
\end{figure}

We test for thermalization by studying the late-time velocity distribution, which is Gaussian in the canonical ensemble, and thus also in the microcanonical ensemble for large $N$ under equivalence of ensembles. In Fig.~\ref{fig:late}, we perform a standard Gaussian test for the velocity distribution of time-average ensembles obtained from evolving some squeezed IC with $\rho_m a = 1/2$ (the most chaotic choice), up to $t\omega = 2\times 10^4$ \XC{(in comparison, pre-thermalization by integrability breaking is studied at $t \sim 10^2$~\cite{Bertini15})}. The result shows a clear deviation from Gaussianity, which persists in the stationary regime and moreover {amplifies} as $N$ increases, barring finite-size effects. In comparison, a modified dynamics which shuffles randomly the velocities every unit time thermalizes far more quickly. Our result does not depend on the IC chosen. Indeed, even for thermally typical ICs, the time-averaged ensembles show visible (although smaller) deviation from Gaussianity ~\footnote{This does not contradict the Liouville theorem, since only a single trajectory is considered. Upon averaging over a few thermal ICs, we obtain a Gaussian as expected.}. Furthermore, the dependence on ICs is unpredictable, due to chaos. 

Nevertheless, we propose a simple description of the late-time ensembles, their phase-space distribution is a \textit{stationary} solution of the tHRE: 
\begin{equation} \partial_x (v \rho) - \partial_x V \partial_p \rho = 0 \,. \label{eq:HRE0} \end{equation}
The idea is quite simple and \XC{similar to discussions in ~\cite{DoyonYoshimura17}}: the late-time ensemble should be void of macroscopic momentum flow on average, which the tHRE calculates to leading order (in a derivative expansion). Since late-time ensemble distributions are usually quite smooth and gently varying, we expect eq.~\eqref{eq:HRE0} to perform well.

To test this idea, we invoke the following fact (see also~\cite{DoyonYoshimura17}): $\rho$ solves eq.~\eqref{eq:HRE0} if and only if the corresponding Fermi factor $\theta$ [see eq.~\eqref{eq:S}] depends only on  $\frac12 p^2 + \int_0^x \left(1-a\int_{p'} \rho(y,p') \right) y \dif y \,.$
This makes it possible to reconstruct the velocity distribution of any solution $\rho$ from its density. We can apply this to the time-averaged density and compare with the true velocity distribution: eq.~\eqref{eq:HRE0} holds if the actual and reconstructed distributions coincide. We performed this test on numerous late-time ensembles, and show two examples in Fig.~\ref{fig:late}. The results are excellent almost everywhere, except that a small discrepancy is observed near sharp central peaks, possibly due to diffusive corrections~\cite{DoyonSpohn17}. Overall, the stationary description is remarkably successful given its simplicity, and suggests a tempting scenario of 
\XC{anomalous thermalization}: the resurrection of tHRE implies that its conserved quantities become again microscopically conserved after time-averaging (for the entropy, this can be seen in Fig.~\ref{fig:S}), and prevents the late-time system from thermalizing further.

\paragraph{Conclusion.} \XC{We studied a classical paradigm of integrability breaking by a trap, which displays dynamical features which are peculiar compared to generic many-body interacting systems. Chaos is suppressed in larger systems, and thorough thermalization takes a prohibitively long time.} The relation with kinetic theory via tHRE is also non-trivial: the latter is valid at short time \textit{and} long time, and breaks down during the intermediate regime.

It would be interesting to explore how far the above findings extend to quantum many-body systems, and it seems reasonable to expect that the three-regime scenario remains valid in regimes where the quantum-mechanical wavelength $\lambda$ is much smaller than $1/\rho$ and $a$ (understood as the scattering length). Otherwise, the eq.~\eqref{eq:timetochaos} is possibly a lower bound: $t_* \geq 1/(C a \rho_m \omega)$. Indeed, in the weakly interacting limit, $a \to \infty$ in 1D, but an infinitely fast advent of chaos is unphysical.  At the same time, quantum coherence may make finite-$N$ systems more resilient to the complexity crisis~\cite{Gutzwiller}. Nevertheless, since the above arguments are general, we expect that the late-time ensemble of a trapped $\delta$ Bose gas still satisfies the corresponding kinetic equation, even in fully quantum regimes.

\paragraph{Acknowledgments.} We are grateful to R. Vasseur for co-developing the algorithm to solve the tHRE numerically and to C. Karrasch for collaborations on previous works.  The authors acknowledge support from a Simons Investigatorship (X. C.), the Chern-Simons Initiative of UC Berkeley and NSF DMR-1507141 (V. B. B.) and the U.S. Department of Energy (DOE), Office of Science, Basic Energy Sciences (BES), under Contract No. DE-AC02-05-CH11231 within the TIMES Program (J. E. Moore).
\bibliography{hardrodbib}


\section{Time scale of complexity crisis equals time to chaos}

In the main text, we show that chaos occurs at a time scale $t_*  \propto 1/(\rho_m a)$. Here, we show that the same scaling relation governs other quantities concerning the initial regime dynamics, including the complexity crisis. This indicates strongly that integrability breaking is the source of both phenomena.

\subsection{Galaxy formation}
\begin{figure*}
	\includegraphics[width=.4\textwidth]{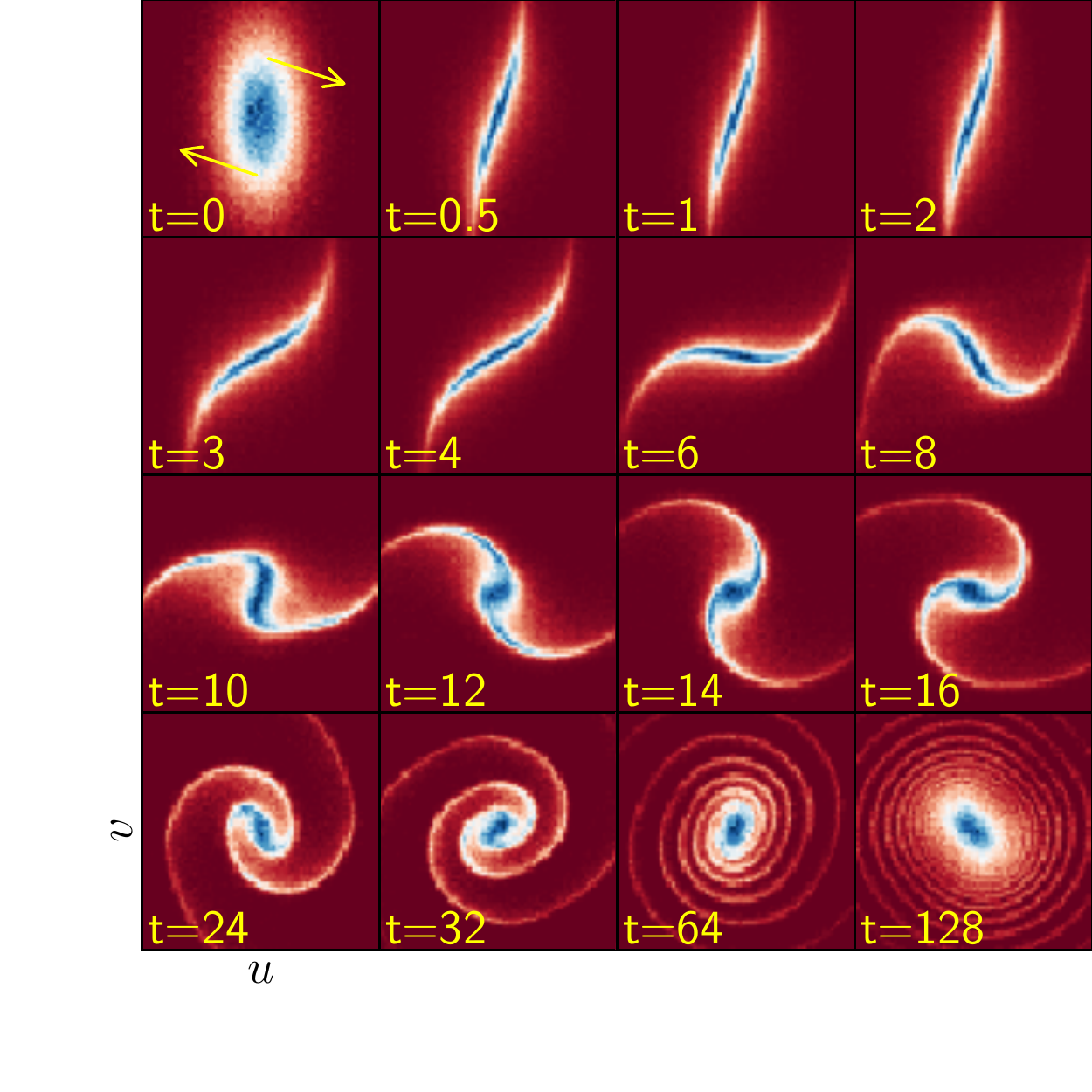}
	\includegraphics[width=.4\textwidth]{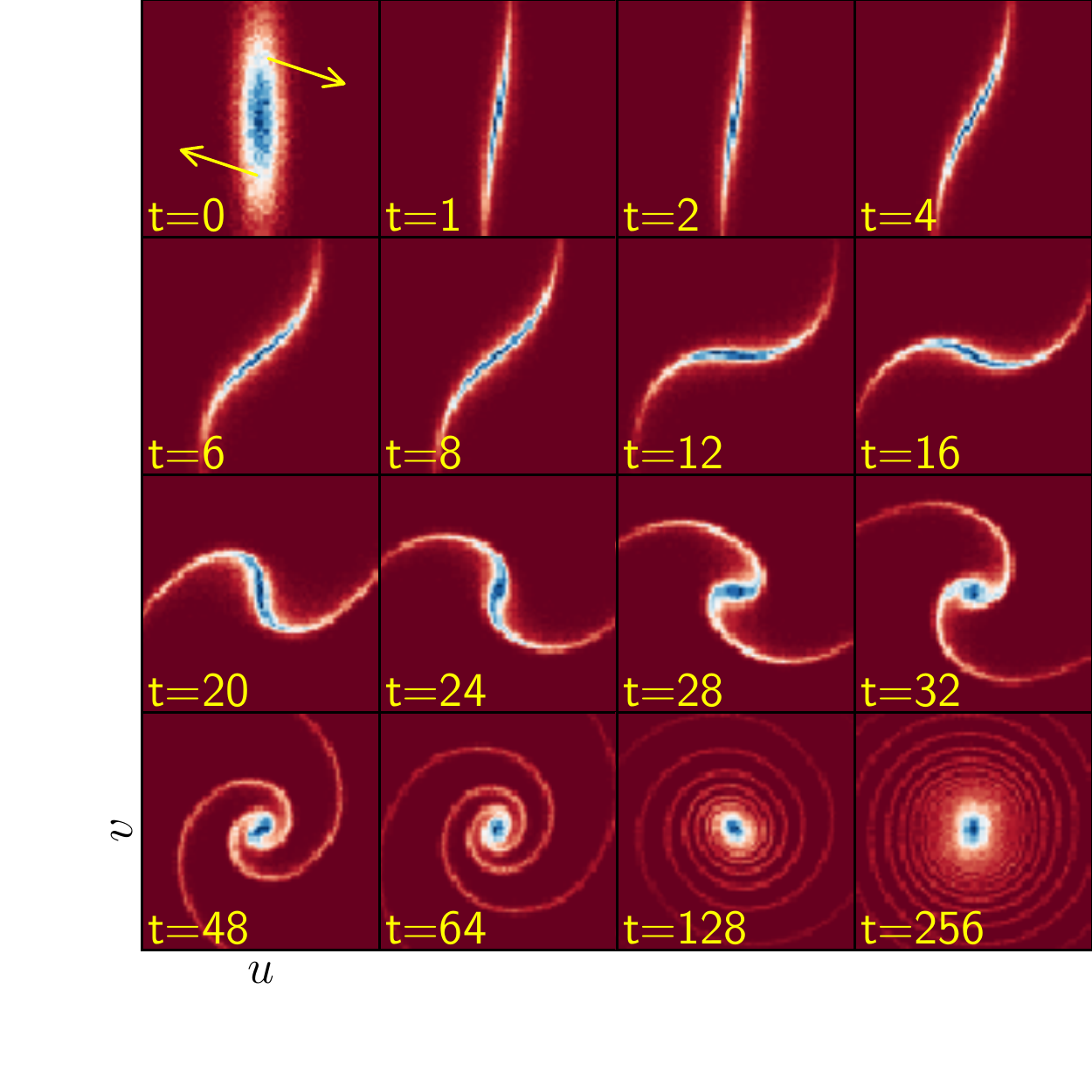}
	\caption{$u$-$v$ phase space distribution of rod dynamics, from two squeezed IC with $\rho_m a = 1/2$ (left) and $\rho_m a = 1/4$ (right). $N = 2048$ and $\sigma_x = 1/2$ for both datasets. In the last plot of each panel, the color bar is non-uniform to highlight the ``galaxy'' structure.
	}\label{fig:milky2}
\end{figure*}

The most direct way of appreciating the complexity crisis is visualizing the ``galaxy'' formation in the $x$-$p$ plane, as done in the main text for different $N$. Here, we compare between different values of $\rho_a m$. We also work in the rotating frame defined by
\begin{equation}
u + \im v := \exp(\im \omega t) (x + \im p)  
\end{equation}
to isolate the effects of hard rod interaction. In Fig.~\ref{fig:milky2}, we compare results from two ICs with different $\rho_m a$ and different $t$, with identical \textit{rescaled} time $t \rho_a m$. Note that, there is no symmetry (microscopic or at the tHRE level) that allows one to vary $\rho_m a$, so we cannot expect the shapes to be exactly identical. Nevertheless, we remark that the angular velocities of the galaxy arm growth are identical in units of the rescaled time $t \rho_m a$, or, equivalently, $$ \text{angular span of arms} \propto \rho_m a  t\,, $$
the left hand side being a measure of the complexity. We conclude that the onset of complexity crisis occurs at the time scale  $t_*  \propto 1/(\rho_m a)$, which is also the time to chaos.

\subsection{Dephasing time: density oscillation damping}

\begin{figure*}
	\includegraphics[width=.4\textwidth,valign=center]{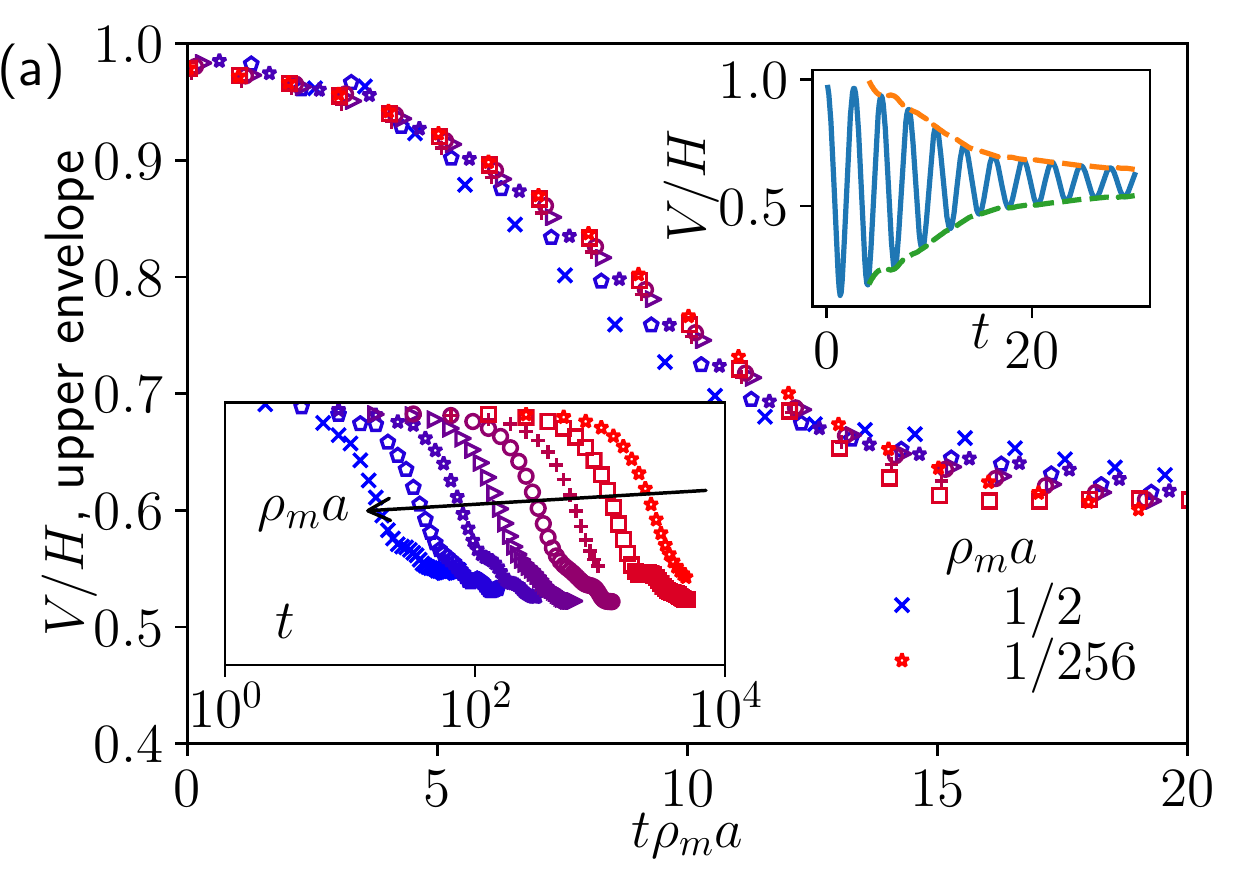}
	\includegraphics[width=.4\textwidth, valign=center]{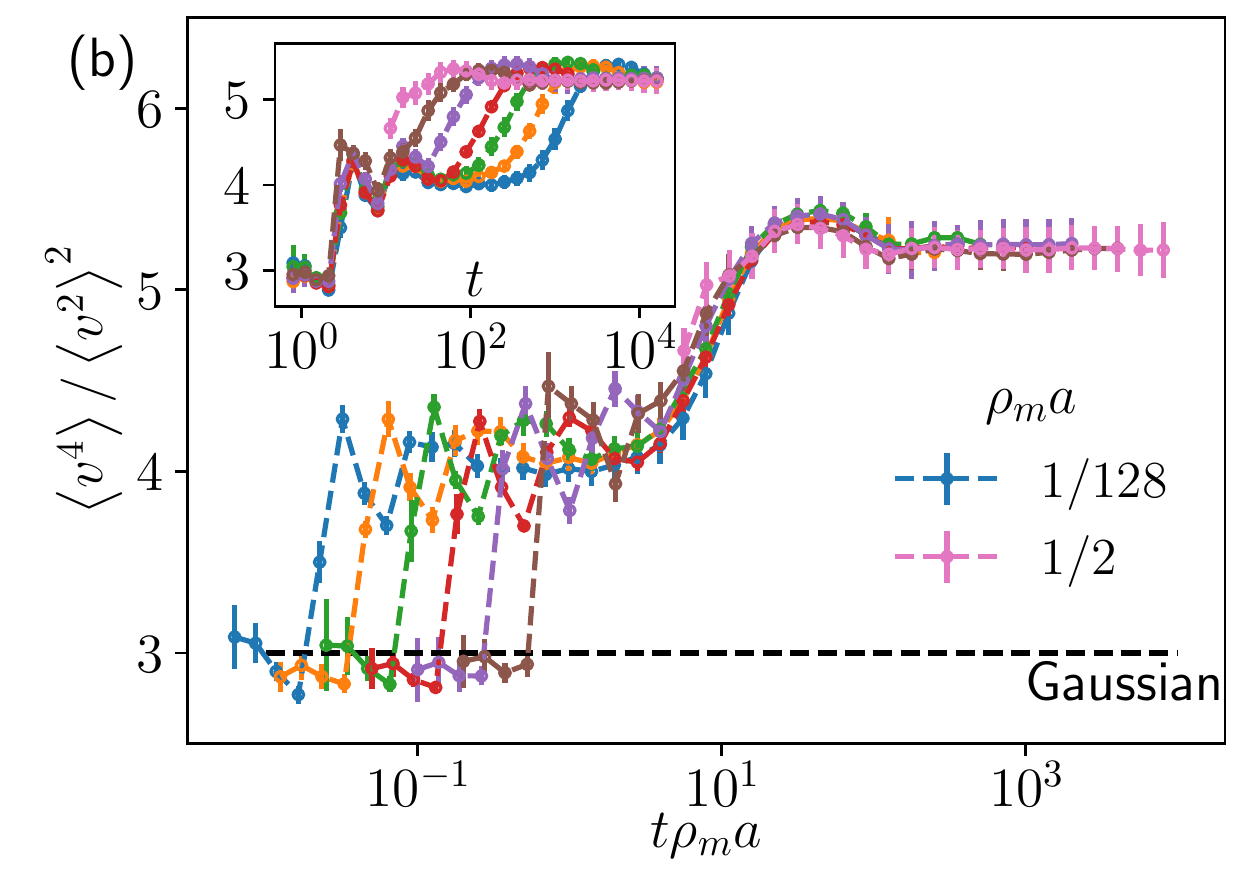}
	\caption{Other manifestations of the scaling law $t \propto 1/(a \rho_m).$ (a) Scaling of the oscillation damping of the ratio $V/H$. Right inset: the evolution of $V/H$ from a squeezed IC with $\rho_m a = 1/2$ is plotted as a solid blue curve. Its envelope is shown in yellow and green dashed lines. Main: IC-averaged upper envelope with different $\rho_m a$. The oscillation damping evolution for different $\rho_m a$ collapse onto one master curve, depending on $t \rho_m a$. Left inset: raw data.
		(b) The Gaussianity measure, same as Fig.~5-a of the main text, with different values of $\rho_m a$. Main plot: data collapse illustrating the  law $t \propto 1/(a \rho_m)$. Inset: raw data.   }\label{fig:VE}
\end{figure*}
As we discussed in the main text, it is the many-body dephasing that generates the complex structure which eventually triggers the complexity crisis. To characterize the dephasing speed more precisely, it is convenient to look at some scalar observable, such as the density as the origin (studied in the main text, Fig.~\ref{fig:compare}). Here, we study a qualitatively similar but numerically more stable quantity: the ratio of the total potential energy \begin{equation}
V := \frac{1}{2} \sum_j \omega x_j^2  \,.
\end{equation}
to the total energy $H$. Being dimensionless, $V/H$ takes value between $0$ and $1$. Starting from a squeezed initial condition, $V/H$ shows a damped oscillation, as shown in Fig.~\ref{fig:VE}0-(a), right inset. The oscillation period is close to $\pi / \omega$ when $\rho_m a$ is small, and decreases slightly as $\rho_m a$ increases: there is no simple scaling law. Yet, the damping speed is proportional  to $\rho_m a$. Indeed, the evolution of the oscillation \textit{envelope},  depends only on the rescaled time $t \rho_m a$, as we show in the data collapse Fig.~\eqref{fig:VE}-(a), main. This shows again that the time of complexity crisis equals that of chaos.

\subsection{Deviation from Maxwellian}
We demonstrate yet another way in which the scaling law $t_* \omega \propto 1/(\rho_m a)$ manifests itself.

In the main text, as evidence for incomplete thermalization, we showed the deviation from Gaussianity of the velocity distribution of the late time ensemble.  Here, we consider the evolution of the time-averaged ensemble (in practice, we average over the time window $[t/2,t]$ and call this the time-averaged ensemble at $t$), for different values of $\rho_m a$. The results are shown in Fig.~\ref{fig:VE}(b). We can again distinguish three regimes:
\begin{enumerate}
	\item  In the initial regime, the evolution does not scale with $\rho_m a$, for a large range of $\rho_m a \lesssim 0.25$: it only depends on $t$. The moment-ratio of velocity distribution attains a plateau after some fast transient. 
	\item  In the chaotic regime, the moment-ratio quits the previous plateau. This regime obeys again the scaling law $t_* \propto 1/(\rho_m a)$; the dynamics depends only on the rescaled time $t \rho_m a$.
	\item In the late time regime, the moment ratio attains another, presumably stationary \textit{non-thermal} value. Remarkably, this is independent of $\rho_m a$: the non-thermalization persists even when the interaction is weak.
\end{enumerate}

\section{Entropy production: integrable case}
In this section, we consider  the hard-rod gas confined by periodic boundary conditions. We will study its entropy production, and compare to the entropy production term of Ref.~\cite{DoyonSpohn17,Boldrighini1997}. We show that in this case, for which integrability is not broken, there is another kind of complexity crisis due to single-particle dephasing (since the period for a soliton to travel around the ring depends on the velocity), which gives rise to an entropy production that is not captured by the entropy production term.  

\subsection{Hydrodynamic scaling}
Let us first review the entropy and entropy production results in Ref.~\cite{DoyonSpohn17},  whose notation we adopt in this section. That is, we let $a$ denote the rod length, $f(x,p)$ the phase space density of the gas and $\rho(x) = \rho[f](x) = \int \dif p f(x,p)$ the spatial density. Then we recall the definition of entropy from eq.~\eqref{eq:S} (of main text):
\begin{equation}
S = - \int \dif x \dif p f(x,p) \ln \frac{f(x,p)}{1-a\rho(x)} \,. 
\end{equation}
Let us discuss its hydrodynamic scaling behavior. By this, we mean setting 
\begin{equation} f(x,p) := f_0(x/N, p/N) / N \end{equation} 
where $N \to \infty$ is the number of particles (if $f_0$ is properly normalized)
$$ \int f(x,p)  \dif x \dif p = N \int f_0(x, p) \dif x \dif p =: N \,. $$
Similarly, $\rho(x) = \rho_0(x/N)$ where $\rho_0$ is a normalized density profile of order unity width. Then, the entropy for $f$ scales as
\begin{equation}
S[f] = N \ln N + N S[f_0] \,.
\end{equation}
In the main text, we rescaled the numerically measured entropy (l.h.s) with $N$ particles and subtracted the extensive behaviors: the quantity plotted in the entropy growth figure is $S / N- \ln N$.

Now we come to the energy production. Eqs. (56) and (57) from Ref.~\cite{DoyonSpohn17} imply the following formula for the total entropy production 
\begin{widetext}
	\begin{equation}
	\dif S / \dif t = \Sigma = \frac14 a^2 \int \dif x  \frac1{1-a\rho(x)} \iint \dif p\dif q \abs{p-q} f(x,p)f(x,q) \left( \partial_x \ln f(x,p) -  \partial_x \ln f(x,q) \right)^2. \label{eq:Sigma}
	\end{equation}
\end{widetext}
This result will be tested below.

The hydrodynamic scaling of the entropy production term $\Sigma$ is trivial:
\begin{equation}
\Sigma[f] = \Sigma[f_0] \,. 
\end{equation}
Therefore, the production of entropy per particle (or per unit length) scales as a $1/N$ correction:
\begin{equation}
\frac{ \dif S }{ N \dif t } =  \Sigma[f] / N \propto 1/N \,.
\end{equation}
This is consistent with the mathematical statement concerning the diffusive correction as a $1/N$ correction in Ref.~\cite{Boldrighini1997} (in which $\epsilon = 1/N$). If such a scaling prevails in some finite time regime, the increase of $S$ should be of order unity independent of the system size. 

\begin{figure}
	\center	\includegraphics[width=.8\columnwidth]{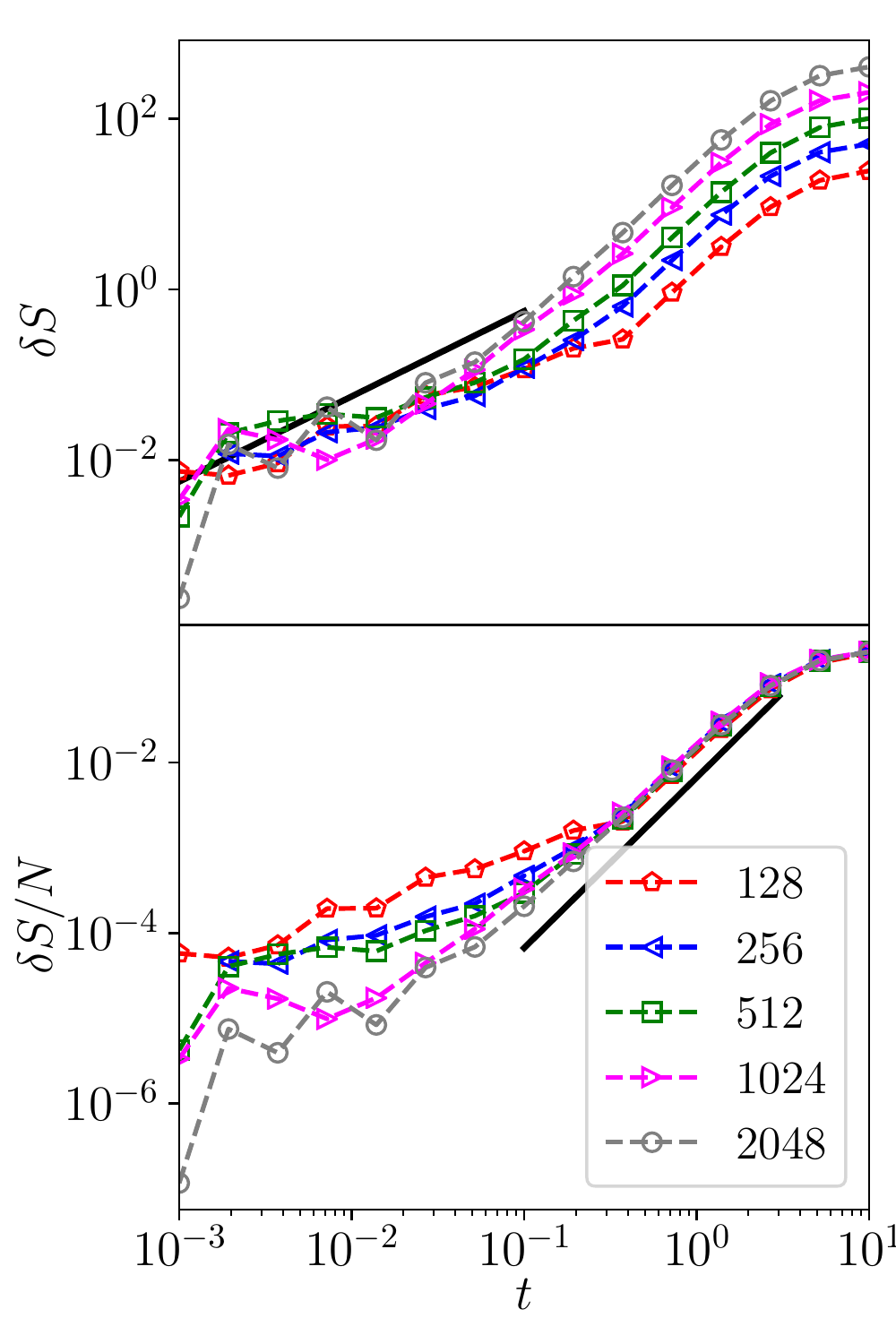}
	\caption{Entropy growth $\delta S = S[f(t)] - S[f(0)]$ in periodically confined hard-rod gas with different system sizes. The parameters are $ c = 1$ and $\sigma = 1$. The upper panel plots the increase of extensive entropy $S$. The black line corresponds to $\delta S = \Sigma[f_0] t = \pi^{3/2} t$, obtained by evaluating eq.~\eqref{eq:Sigma} with the profile eq.~\eqref{eq:f0sin}. The lower panel plots increase of the entropy \textit{density} $S/N$. The black line indicates a power law $\propto t^{2}$.  }\label{fig:flatS}
\end{figure}

\subsection{Numerical study}
The above considerations can be used to test numerically whether the entropy production formula is useful at finite time. For this purpose, we consider the hard rod gas with $N$ particles of rod length $a = 1$, in a periodic ring of length $2N$. This setting introduces a confinement while preserving integrability, so can be instructively compared to the trapped hard rod gas, which breaks integrability.

The (random) initial configuration is generated  such that $q_j = x_j - j, j = 1,\dots,N$ are the ordered statistics of $N$ i.i.d random variables in $[0, N]$, and $p_j = c N \sin(2 \pi q_j / N)  + \sigma N G_k$, where $G_1,\dots,G_N$ are i.i.d standard Gaussians independent of $q_j$. This corresponds to a family of initial profiles which scale nicely as follows: 
\begin{align}
& f(x,p) = f_0(x/N,p/N) / N \nonumber \\
& f_0(x,p) = \frac12 \frac{1}{\sqrt{2 \pi} \sigma} \exp\left( - \frac{(p - c \sin (\pi x))^2}{2 \sigma^2 }   \right) \,. \label{eq:f0sin}
\end{align}
Note that the spatial density $\rho = 1/2$ is constant, but the position is correlated with the velocity, resulting in a non-trivial hydrodynamic profile.  Note also that we always scale the velocity and space together, so that it takes typically $t = 1$ time for a soliton to travel around the ring once.

\begin{figure*}	
	\center	\includegraphics[width=.8\columnwidth]{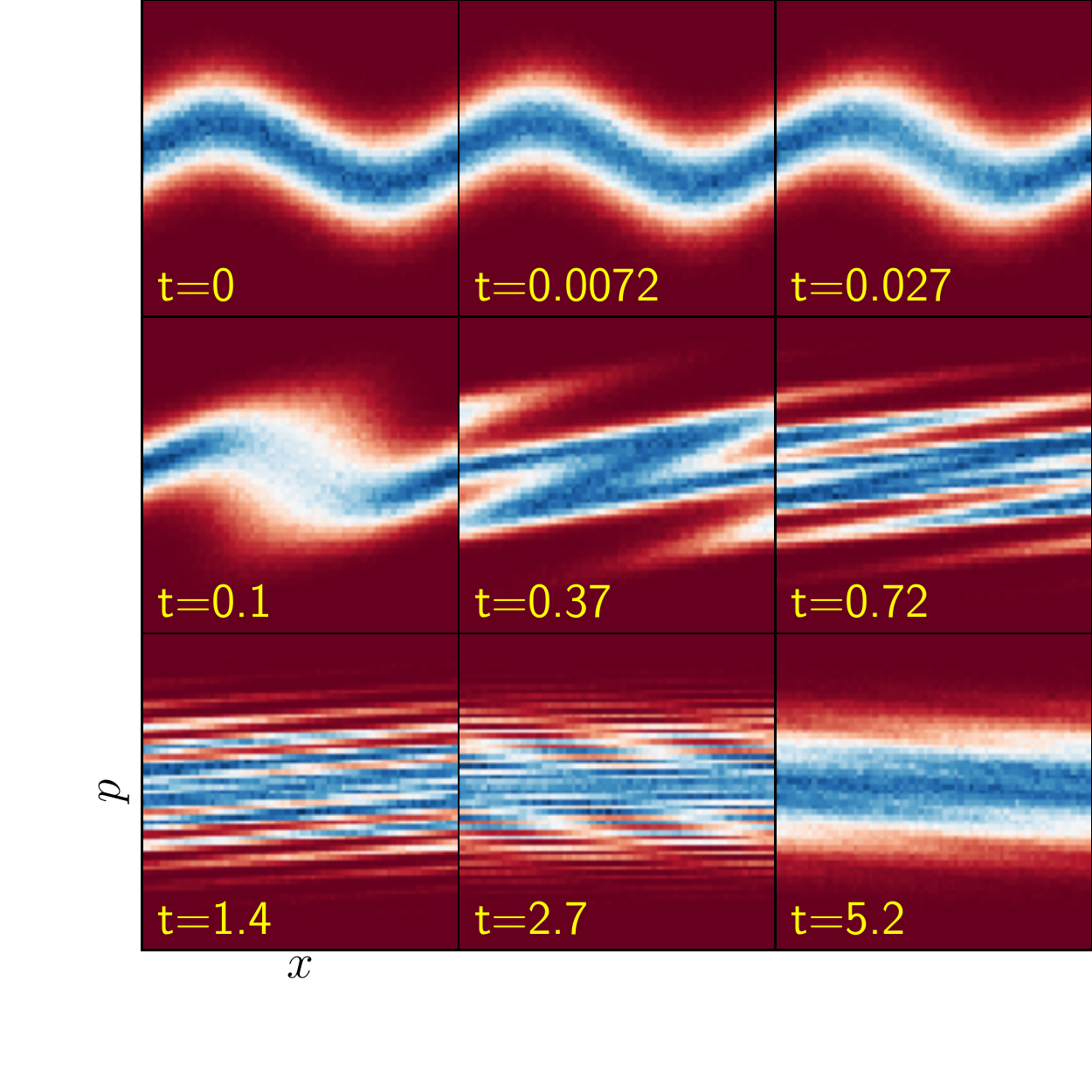} 
	\includegraphics[width=.8\columnwidth]{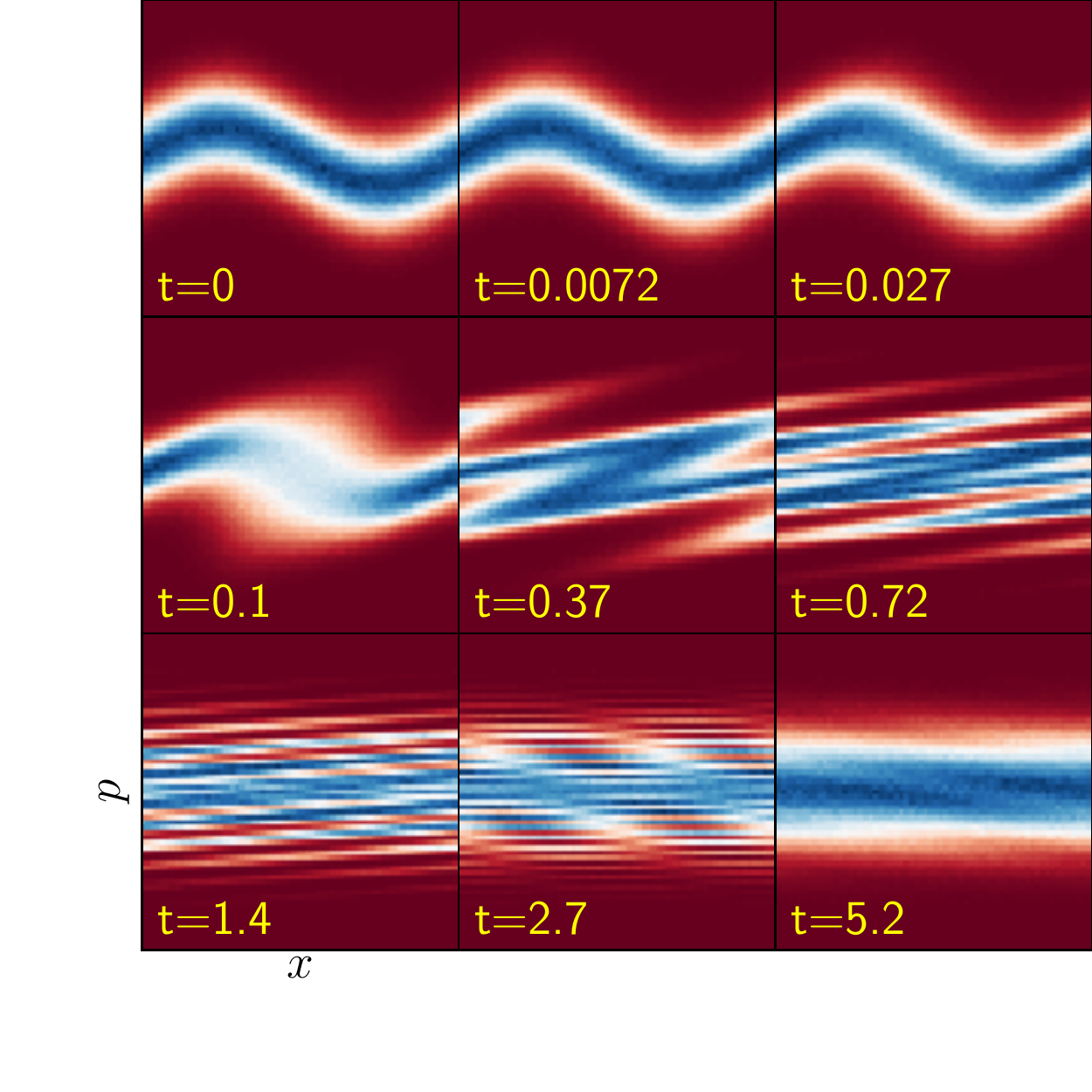}
	\caption{Phase space distribution from numerical simulation, with the initial condition of eq.~\eqref{eq:f0sin}, for $N = 128$ (left) and $N = 2048$ (right). The distributions are nearly identical, even during the integrable complexity crisis (ICC).}\label{fig:milkring}
\end{figure*}

We simulated exactly the evolution of this initial condition and measured the entropy by numerically integrating eq.~\eqref{eq:S}. Some phase space distributions used to calculate the entropies are shown in Fig.~\eqref{fig:milkring}. To compute the entropy, some binning is necessary. We shall use the same mesh (of size $128 \times 128$) for all measures.  The results are plotted in Fig.~\ref{fig:flatS}. We do observe an entropy increase for all finite but moderately large sizes $N = 10^2 \sim 10^3$ that are studied. We observe three regimes:

In the very short time regime, $t \lesssim 10^{-2}$, the entropy increase $\delta S = S\vert_{0}^{t}$ is independent of $N$, and is consistent with the prediction of eq.~\eqref{eq:Sigma}, despite significant statistical noise: going to shorter times is numerically hard, but the main theorem of Ref.~\cite{Boldrighini1997} will guarantee the result. Note that in this regime, the phase space distribution is essentially unchanged compared to the initial profile, see Fig.~\ref{fig:milkring} (the first row of each panel). 

In the unit time regime $t \sim 1$, the phase space distribution changes considerably. the entropy increase is proportional to $N$.  Such a scaling behavior is \textit{qualitatively} incompatible with that of the entropy production formula, thus it is not useful to further compare them quantitatively by evaluating eq.~\eqref{eq:Sigma} (the result of such an attempt is a failure by orders of magnitudes). The unit time regime is also characterized by a supra-linear entropy growth; the power law observed in Fig.~\ref{fig:milkring} is slightly smaller than quadratic.

Finally, there is a saturation regime $t\geq 10$, in which the entropy ceases to grow. We expect that in the long time limit, the positions of the rods/solitons will be decoupled from their velocities, and the system falls into a generalized Gibbs ensemble (GEE) characterized by its velocity distribution: $f(x,p) \to \rho \, \text{PDF}(p)$ where $\text{PDF}(p)$ is the conserved velocity distribution.
The cross-over mechanism from the unit time to the saturation regime is an \textit{integrable} complexity crisis: in Fig.~\ref{fig:milkring}, the phase space distribution also becomes increasingly complex, due to single-particle dephasing: the period depends on the velocity. Such a phenomenon takes place in any integrable model which is not maximally super-integrable, e.g., a non-interacting gas in a non-harmonic trap. Any fixed test function/binning will fail to resolve the complex structure at some point. Thus a partial thermalization towards to the GGE occurs effectively; this is illustrated by Fig.~\ref{fig:milkring} (last row). 

\subsection{Discussion}
Let us compare the integrable complexity crisis (ICC) with the trapped one (TCC). The different mechanisms (single-particle vs. many-body {dephasing}, integrable vs. non-integrable {dynamics}) lead to the following conclusions:
\begin{itemize}
	\item  The ICC can occur even in the non-interacting $\rho a \to 0$ limit. In that limit TCC does not occur (indeed, it becomes a maximally super-integrable model).
	\item  The ICC is not suppressed by increasing the number of particles, as is the case of the TCC (compare Fig.~3-a of main text with Fig.~\ref{fig:milkring}).
\end{itemize}
Therefore, the trapped hard rod case is the first example where the effect of the \textit{integrable} diffusive correction term predicted in Refs.~\cite{Boldrighini1997,DoyonSpohn17} (and earlier by Spohn) is numerically demonstrated to be useful in describing entropy growth beyond the very short time regime. 

While the absence of single-particle dephasing is crucial for the non-integrable trapped system to be ``more integrable'' in a special sense, the above numerical finding raises an intriguing question regarding the roles of chaos and integrability breaking. We know from above that the time scale of the TCC is identical to that of chaos, that is, $t_* = C / (\rho_m a)$: this suggests some relation between them. If this is the case, how can we explain the success of the integrable diffusive correction term? We believe the answer lies in the local equilibrium assumption of the mathematical result~\cite{Boldrighini1997}. In the trapped case, the chaos leads to a local equilibration which enables the diffusive correction eq.~\eqref{eq:Sigma} to perform well. In contrast, such a mechanism is completely absent in the integrable ring system, which has a completely different ``equilibration'' mechanism.

\end{document}